\documentstyle[12pt]{article}
\begin{document}
\baselineskip=20pt

\title{\bf Bounds on the cosmological abundance of Primordial Black Holes\\
from diffuse sky brightness: single mass spectra.}

\author{P.S.Cust\'odio and J.E.Horvath\\
\it Instituto Astron\^omico e Geof\'\i sico\\
Universidade de S\~ao Paulo\\
Av. M. St\'efano 4200 - Agua Funda\\
04301-904 S\~ao Paulo SP Brazil\\}
\maketitle
\pagestyle{empty}
\vskip5mm

\noindent{\bf Abstract}

We constrain the mass abundance of unclustered  primordial black
holes (PBHs), formed with a simple mass distribution and subject to the
Hawking evaporation and particle absorption from the environment.
Since the radiative flux is proportional to the numerical density,
an upper bound is obtained by comparing the calculated and
observed diffuse background values, (similarly to the Olbers
paradox in which point sources are considered) for finite
bandwidths. For a significative range of formation redshifts the
bounds are better than several values obtained by other arguments
$\Omega_{pbh} \, \leq \, 10^{-10}$; and they apply to PBHs which are 
evaporating today.

\vfill\eject
\section{Introduction}

A variety of observations make a compelling case for the existence
of dark matter in unknown form(s). Among the
discussed possibilities we may mention axions, WIMPS, cosmic
strings, brown dwarfs and Primordial Black Holes, see
\cite{[PEA98]}. It has been longely recognized that the abundance
of evaporating PBHs \cite{[HAW75]} may be constrained by a variety
of methods . For example, Carr and Mac Gibbon (see \cite{[CMC98]}
and references therein) used the gamma-ray diffuse background to
establish constraints. Similarly, Liddle and Green \cite{[LG98]}
reviewed a wide variety of methods to study a possible PBH
population. Generally speaking, the methods dealing with the
evaporation of PBHs apply to the final phase of these objects
only, and therefore do not make full use of the radiative thermal
flux at larger wavelengths predicted by the Hawking process of
quantum evaporation \cite{[CMC98]},\cite{[LG98]}.

The formation of PBHs
from primordial fluctuations was first studied by Carr
\cite{[CARR75]}. In that work he established that large, gaussian
primordial fluctuations from a scale-free primordial spectrum may
be responsible for a power-law PBH mass function. A new twist to
the problem has been added by the recent work of Niemeyer and
Jedamzik \cite{[NJ98]},\cite{[NJ99]} who used the concepts of
critical phenomena to study the initial mass function for Gaussian
and "mexican hat" initial fluctuation spectra. They have argued
that in these cases PBHs may form with masses  below the
horizon mass, and found a mass function which is not a pure
power-law. Black holes may form at all epochs 
(see Yokoyama \cite{[YOK98]}, and Kawasaki and Yanagida \cite{[KY98]} for recent
work on PBH formation from the collapse of large density perturbations 
in double-inflation and supergravity models) and
since there is considerable uncertainty about the actual processes
that effectively formed PBHs, we seek here a general method to
constrain the population which may be used independently of the
specific PBH formation mechanism(s) (see for instance, 
Hansen et. al., and Bousso \cite{[BOU98]} for a discussion of other possibilities). 
We shall evaluate a simple
(but useful) model consisting in a single-mass scale, which may be
associated to the typical mass-scale of any given mass function
showing such a peak (this situation is very well satisfied by models in 
which large perturbations collapse to form PBHs).

The simplest reexamination of PBH physics has shown that, as a result of
particle absorption from the expanding background, there is a
maximum value of the mass (termed "critical mass" in
\cite{[CH98]}) splitting an evaporating (or subcritical) from a
non-evaporating (or supercritical) subset of a general mass
distribution. The critical mass separating these regimes for 
each redshift is therefore a useful boundary to discuss the 
physics of the PBHs and will be defined in the next section . A
detailed quantitative discussion of these issues has been given in
Refs.\cite{[CH98]},\cite{[CH99]}. Due to the absorption of
particles from the thermal background, PBH evaporation (and
therefore some limits to their abundance) have to be reexamined.
Actually we shall see that the existence of a regime of $\sim$
constant mass in the life of PBHs is very important for a full
evaluation of their survival. Previous studies \cite{[BAR91]} have
not considered the regime of quasi-constant mass and thus need to
be updated to account for the evaporation at each evaporation
scale labeled by the redshift.

\section{Formalism}

In order to simplify our evaluations, let us consider a
homogeneous and isotropic cosmological distribution of PBHs. The
radiative thermal flux from an evaporating black hole, as measured by an
observer at $z=0$ is given by
$F(M)={L(M)\over{4\pi{D_{L}^{2}(z)}}}$, where
$D_{L}(z)={cz\over{H_{0}}}\left[1+{z(1-q_{0})\over{
{\sqrt{1+2q_{0}z}+1+zq_{0}}}}\right]$ and $q_{0}$ is the
deceleration factor (see, for instance, \cite{[PEA98]}).

The luminosity function $L(M)$ is given by
$L(M)={L_{0}}{\left(M_{Haw}/M\right)}^{2}$, with
$L_{0}\sim{2.6 \times {10}^{16}}erg \, {s}^{-1}$ and it is simply the
Stefan-Boltzmann's law applied to an evaporating black hole. We have scaled
the mass to a "natural" constant $M_{Haw}\, = \, {10}^{15} g$ 
(hereafter the Hawking mass), since this is the mass of a 
PBH which is completing 
its evaporation today. Thus we write $\mu_{*}\, \equiv \,
({M_{*}\over{M_{Haw}}})$ and so on along this work.

The subscript $"*"$ will be used for the initial values of any
quantity. Moreover, we shall restrict our considerations to
semiclassical PBHs, i.e. those that satisfy $M  \gg {M_{pl}}$, 
$M_{pl}$ being the Planck mass.

We define $d\mu \xi(\mu,z)$ as the
numerical abundance of PBHs $dn$ at redshift
$z$ with masses between $\mu$ and $\mu+d\mu$ per unit horizon volume;
then $\xi (\mu,z)$
denotes the number of PBHs by horizon volume and by mass interval and
 ${\Delta}V_{hor}(z)d\mu\xi(\mu,z)$ is the number of PBHs contained
within this volume. An upper limit on the mass density of
radiating PBHs is obtained from the requirement
$\delta{F}_{total}(PBHs) \, < \, \delta{F}_{back}$, with $\delta F_{back}$ the 
value of the measured background flux in a given frequency interval. 

To begin the calculation we assume a dilute PBH gas, i.e. a negligible collision
term (PBH-PBH) between these objects. In other words we assume
that the cosmic expansion rate is much larger than thePBH
collision rate. Our aim will be to obtain windows of allowed
abundances, for which we limit the PBH mass abundance in function
of initial mass versus formation redshift.

As discussed in Ref.\cite{[CH98]} , the minimal formalism that
describes these features is given by the set of equations

$$H^{2}={{{8\pi}G}\over{3}}{\varrho_{total}}-{K\over{R}^{2}}+{\Lambda\over{3}}
\eqno(1)$$

$${\dot{\mu}}=-{A(\mu)\over{{\mu}^{2}}}+{27\pi{G}^{2}\over{{c}^{3}}}{\mu}^{2}
{\varrho_{rad}(t)}\eqno(2)$$

$$4H{\varrho_{rad}}+3H{\varrho_{pbh}}+{\dot{\varrho}}_{pbh}
+{\dot{\varrho}}_{rad}={{\dot{Q}(t)}\over{{R(t)}^{3}{c}^{2}}}\eqno(3)$$

$${Df_{pbh}\over{Dt}}={\partial{f_{pbh}}\over{\partial{t}}}+ {d\mu\over{dt}}
{\partial{f_{pbh}}\over{\partial{\mu}}}
-H{\vec{\beta}}_{pec}.{\partial{f_{pbh}}\over{\partial{\vec{\beta}}_{pec}}}
 = 0 \eqno(4)$$

Where $\varrho_{total}$ denotes the total density (radiation plus
PBHs), $\dot{Q}(t)$ gives the total heat input from the evaporation from
the subcritical PBH population 
($\dot{Q} (t) \, = \, \int dm \, {\dot m} N(m,t)$), $R(t)$ is the usual scale factor,
$f_{pbh}=f_{pbh}(\mu,\beta,t)$ is the distribution function that
describes the cosmological evolution of PBHs, their numerical
abundances and higher momenta and ${\vec{\beta}}$ denotes the
peculiar velocity of the PBH as given by
$\vec{\beta}=({\vec{v}}/c)$. In eq.(2)
$A(\mu)\sim{10}^{26}g^{3}{s}^{-1}h(\mu)$ with
$h(\mu>10^{2})\sim{1}$ and $h(\mu<<1)\sim{100}$, describing the 
degrees of freedom of the emitted particles according to the Standard
Model.

The critical mass value (as introduced in \cite{[CH98]})
separating the evaporating (subcritical) from the non-evaporating
(supercritical) regimes is obtained from the condition
$\dot{\mu}=0$ and reads

$$M_{c}(T)\sim{7.3\times{10}^{25}g
{\biggl[{\varrho_{rad}(t_{0})\over{\varrho_{rad}(t)}}\biggr]}^{1/4}}
\sim{7.3\times{10}^{25}g\over{(T/T_{0})}}  \eqno(5)$$

with $T_{0} \sim {2.7 K}$ the present value of the CMBR
temperature, and the other symbols have standard meaning. In terms
of the Hawking mass we write eq.(5) as
$\mu_{c}(T)\sim{10^{11}\over{(T/T_{0})}}$.

Once the individual fluxes from the evaporating subpopulation 
are calculated, the total flux can be
obtained by integration

$$\delta{F}_{pbh}={1\over{4\pi}}
{\int_{0}^{z_{f}}{dz\over{{D_{L}}^{2}(z)}}\left[\Delta{V_{hor}(z)}
\over{\Delta{z}}\right]}
{\int_{\theta \mu_{pl}}^{\mu_{c}(z)}d\mu \xi(\mu,z)L(\mu)}  \eqno(6)$$

where $z_{f}$ denotes the formation redshift and $\theta$ is a dimensionless number 
$O(1)$. Primordial Black
Holes above $\mu_{c}(z)$ do not contribute to the radiation since
for them the absorption term is dominant by definition. As it
stands from eq.(6), a suitable evaluation of $\delta F_{pbh}$
requires integration over the redshift associated to the
expansion. This feature is depicted in Fig.1 which shows the
light-cone and the received radiation emitted from these objects.
All those PBHs with $\mu_{*} < 1$, the Hawking mass, formed at
$z_{f} \gg 1$ had evaporated at $z_{evap}(\mu_{*})>0$.


Generally speaking the functional form of the amplitude
$\xi(\mu,z)$ will depend on the physics of the process that forms
the PBHs and can be quite complicated. A reasonable {\it ansatz} is to
factorize the usual volumetric dilution from the mass dependence.
Thus, the function $\xi (\mu,z)$ for the original \cite{[CARR75]}
Carr mass function is chosen to have the form

$${\xi(\mu,z)}={\xi_{c}(\mu,z)\over{V_{hor}(z)}}
={\xi_{0}}{{{\mu}^{-n(z)}}\over{V_{hor}(z)}}  \eqno(7)$$

and, for instance, 
the critical Niemeyer-Jedamzik is {\it not} a power-law and we refer 
to the references \cite{[NJ98]}, \cite{[NJ99]} for a full account 
of their work. A large class of initial mass functions are nonetheless 
included in the {\it ansatz} class.   

The single-mass scale mass function to be studied in this paper 
is simply

$$\xi(\mu,z)={\xi_{0}}{\delta(\mu-\mu_{*})\over{V_{hor}(z)}}\eqno(8)$$

with the horizon volume given by
$V_{hor}(z)={4\pi{r_{0}}^{3}\over{3{(1+z)}^{3}}}$, the
numerator denotes the present value for the particle horizon
volume and the numerical density in black holes is normalized to
the present value when $z=0$. This normalization factor will be
canceled by the same factor present in $\Delta{V_{hor}(z)}$ in
eq.(6).The subscript $"{c}"$ denotes comoving values and $\xi_{0}$
is the amplitude, interpreted as PBH number by mass interval, i.e.
$PBHs\times{g^{-1}}$. This work will evaluate mass constraints
using the mass function of eq.(8), considered as a basic model. 
Other examples like the
critical collapse spectrum \cite{[NJ98]}, \cite{[NJ99]} discussed
recently, or Carr's power-law spectrum can be worked out, 
although the evaluations are quite involved. However, it is important 
to stress that all power spectra that produce a substantial number 
of PBHs present sharp peaks, and therefore look pretty much like 
delta-functions \cite{[com]}.

To proceed with the evaluation of the bounds we must 
make a connection between the mass function amplitude and 
the solutions of eqs.(1-4)  
in the following way: the number density of PBHs is given by

$$n_{PBH}(z)=\int{dM \xi(\mu,z)}  \eqno(9)$$

On the other hand, we may be able to evaluate $n_{PBH}(z)$ using
the kinetic formalism  (see \cite{[CH98]}) as given by

$$n_{PBH}(\mu,z)={g_{*}\over{{(2\pi)}^{3}}}\int{{d}^{3}pf_{pbh}(\beta,\mu,z)}
\eqno(10)$$

with $p = m c \beta$ is the non-relativistic momentum.

Some algebraic manipulations enable to cast eq.(9) in the form

$$n_{PBH}(z)={g_{*}{c}^{3}{(M_{Haw})}^{4}\over{2{\pi}^{2}}}
\int_{\theta \mu_{pl}}^{{\mu}_{c}(z)}{d\mu}{\mu}^{3}{\int_{0}
^{1}{d\beta}{\beta}^{2}f_{pbh}(\beta,\mu,z)}  \eqno(11)$$

Here, $g_{*}$ is the statistical weight and we integrate up to the
critical mass $\mu_{c}(z)$ since above it the evaporation does not
take place. Note that for $\mu\geq{\mu_{c}(z)}$ the third term of
eq.(4) changes sign and the lower limit $\theta \mu_{pl}$ precludes
contribution of quantum PBHs near the Planck scale. Now, we may
rewrite $n_{PBH}(z)$ using the same normalization factor as

$$n_{PBH}(z)={M_{Haw}}\int_{\theta \mu_{pl}}^{{\mu}_{c}(z)}{d\mu}\xi(\mu,z)
\eqno(12)$$

and by comparing eqs.(11) and (12) we obtain

$$\xi(\mu,z)={g_{*}{c}^{3}{(M_{Haw})}^{3}{\mu}^{3}\over{2{\pi}^{2}}}
\int_{0}^{1}{d\beta}{\beta}^{2}f_{pbh}(\beta,\mu,z)
 \eqno(13)$$

Clearly the microphysical framework is a powerful tool, since it
can describe all the kinematical effects as explicitly displayed
in the limits of the integrals eqs.(11) and (12) above.
A complete analysis of the solutions is beyond the scope of the 
present work and will be
treated elsewhere. Here, we will analyze only the radiation
emitted by these objects.

Our general picture is now formally complete, in the sense that
solving the formalism for $f_{pbh} (\beta,\mu,z)$ we would be able
to set upper limits to the abundance of PBHs using the thermal
emission of any mass spectrum. Since the work done in the
literature mainly addresses the final explosive phases of these
objects, which contributes to the gamma-ray band and to the cosmic
ray flux,(see for example \cite{[CMC98]} and references therein), we
may view this work as an attempt to generalize those results.

\section{ Application to a single-scale mass function and
comparison with other methods.}

To proceed we must now evaluate the last integral found in eq.(6),
for the delta mass-function case where
$\xi{(\mu,z)}\propto{\delta(\mu-\mu_{*})}$. The limits on
$\xi_{0}$ will thus apply to an unclustered isotropic distribution of
PBHs that are subdominant for the total mass balance of the
universe. These hypothesis are in fact not very restrictive since for low
peculiar velocities and the masses considered, the time scale for
clustering is much larger than the Hubble time ${H_{0}}^{-1}$.

As explained above, we
will assume that at $z_{f}$ the mass is concentrated at one
discrete value. Figure 2 shows the two important regimes to be
considered here: $\mu_{*}\geq{1}$ or $\mu_{*}\leq{1}$ in terms of
the formation redshift and the critical mass $\mu_{c}(z)$. To 
visualize the differences we shall discuss them separately.


\bigskip

\noindent
{\bf First regime: $\mu_{*} \geq {1}$}

In this case, all PBHs formed at $z_{f}$ are still evaporating today.
Here, $M_{Haw}\sim{10}^{15}g$ is the Hawking mass, i.e. it is the
mass that completely evaporates today. We recall that the time
scale for evaporation depends only of $\mu_{*}$ and it is given by
$t_{evap}=f(\Omega_{0}){H_{0}}^{-1}{(\mu_{*})}^{3}$. We may split
this regime in two sub-cases depending on wether $z_{f}$ the PBH
mass is larger or smaller than the critical mass
($\mu(z_{f})\geq{\mu_{c}(z_{f})}$ or
$\mu(z_{f})\leq{\mu_{c}(z_{f})}$) at $z_{f}$. In the first subcase, the PBHs
the onset of evaporation will be delayed until they cross the
critical mass curve, as will be described later. Now, we will
study the subcase $\mu_{*}(z_{f})\leq{\mu_{c}(z_{f})}$ which
consists of initially subcritical PBHs.

\subsection{Subcase \, $\mu_{*}(z_{f})\leq{\mu_{c}(z_{f})}$}

These PBHs will start to evaporate immediately after their
formation. The evolution of this spectrum with $z$ will be given
by

$$\xi(\mu(z),z\leq{z_{f}})={\xi_{0}}
{\biggl[{\delta(\mu(z)-\mu_{*}(z))\over{V_{hor}(z)}}\biggr]}
\eqno(14)$$

where $\mu(z)$ is given by the
$\mu(z)=\mu_{*}F(\mu_{*},z_{ini},z)$, and
the evaporation
function $F(\mu_{*},z_{ini},z)$ is 

$$F(\mu_{*},z_{ini},z)\sim{\biggl[{1
+{\biggl({1\over{{\mu_{*}}}}\biggr)}^{3}
\int_{z_{ini}}^{z}dz{(1+z)}^{-5/2}}\biggr]}^{1/3}\eqno(15)$$

which can be approximated by
$({1-{2\over{3 {\mu_{*}}^{3}{(1+z)}^{3/2}}}})^{1/3} \, \sim \, 1$, for $z_{ini}>>1$.

This form is valid as long as $q_{0}=1/2$, because we have used
that the horizon volume at $z$ is related to the horizon volume at
$z_{f}$ by $V_{hor}(z)={\biggl({1+z_{f}\over{1+z}}\biggr)}^{3}
V_{hor}(z_{f})$). We do not consider other cases for $q_{0}$ in
this article, but we checked that the dependence of the results
with $q_{0}$ is very mild.

Substituting these relations into eq.(6), we evaluate the
radiation received today $(z=0)$ from all these PBHs yielding

$$\delta{F_{pbh}}={3L_{0}M_{Haw}\xi_{0}\over{4\pi}}
\int_{\epsilon_{\odot}}^{z_{ini}}dz{{[\mu_{*}F(\mu_{*},z)]}^{-2}
\over{{D_{L}}^{2}(z)(1+z)}}\eqno(16)$$

Assuming that PBHs have black body spectra, we can apply Wien's
law to the mass-temperature relation of black holes:
$T(\mu)\sim {2 \times 10^{11}} K /{(\mu)}$ in order to obtain
an associated wavelength of the maximum
$\lambda(\mu_{*})T(\mu_{*})=0.29 \, cm \, K$. Since we are
considering the radiation emitted by PBHs at cosmological
distances, we must correct for the expansion of the received
wavelength. We know that these relation would be satisfied by
$\lambda_{obs}(z)=\lambda_{em}(z^{\prime}){(1+z^{\prime})
\over{(1+z)}}$. Substituting into the mass-temperature relation 

$$\lambda_{obs}(z=0,\mu(z^{\prime}))={\lambda_{0}}{\mu(z^{\prime})}
(1+z^{\prime})\eqno(17)$$

And considering that at any time the mass is given by
$\mu(z^{\prime}) ={\mu_{*}}F(\mu_{*},z_{ini},z^{\prime})$ we
obtain

$$\lambda_{obs}(z=0,\mu_{*})={\lambda_{0}}{\mu_{*}}
F(\mu_{*},z_{ini},z^{\prime})(1+z^{\prime})\eqno(18)$$

The prefactor in front of $(1+z^{\prime})$ is the emitted
wavelength when the initial PBH with mass $\mu_{*}$ had the mass
$\mu(z)$ at the correspondent temperature, and
$\lambda_{0}\sim{1.5\times{10}^{-12}}cm$ is a reference scale.
Then, we recover ${\mu_{*}}F(\mu_{*},z_{ini},z^{\prime})=
{\lambda_{obs}(z=0,\mu_{*})\over{\lambda_{0}(1+z^{\prime})}}$. By
its very definition $F(\mu_{*},z_{ini},z^{\prime}=z_{ini})=1$, and
thus

$${\lambda_{0}\over{\lambda_{obs}(z=0,\mu_{*})}}=
{1\over{\mu_{*}(1+z_{ini})}}\eqno(19)$$

After all these manipulations the total flux obtained is given by

$$\delta{F_{pbh}}={3L_{0}M_{Haw}\xi_{0}\over{4\pi}}
{\biggl[{\mu_{*}(1+z_{ini})}\biggr]}^{-2}\int_{\epsilon_{\odot}}^{z_{ini}}
dz{(1+z)\over{{D_{L}}^{2}(z)}}\eqno(20)$$

\bigskip

where $z_{ini}=z_{f}$ if $\mu_{*}\leq{\mu_{c}(z_{f})}$ or
$z_{ini}=z_{cross}(\mu_{*})$ if $\mu_{*}\geq {\mu_{c}(z_{f})}$.

\bigskip
An inspection of eq.(20) shows that there is a potential mathematical 
divergence at the lower limit of the integral. Therefore we have 
integrated formally the expression up to a minimum parameter 
$\epsilon_{\odot}$, which acts as a cutoff. The integral 
$I(\epsilon_{\odot},z_{f})=\int_{\epsilon_{\odot}}^{z_{f}}
dz{(1+z)\over{{D^{2}_{L}(z)}}}$ admits the expansion

$$I(\epsilon_{\odot},z_{f})\sim{Y(q_{0}){(H_{0}/c)}^{2}
\biggl[{1\over{\epsilon_{\odot}}} +ln(z_{f}/\epsilon_{\odot})+...
\biggl]} \eqno(21)$$

with  $Y(q_{0})\sim{O(1)}$ and $\epsilon_{\odot}<<1$. The question 
arises about how this $\epsilon_{\odot}$ affects the final results. We shall 
show below that a careful consideration of the different regimes of the 
PBHs makes unnecessary the imposition of any $\epsilon_{\odot}$, 
and in practice there is no physical divergence of the quantities, contrary to 
the naive expectation from eq.(20).

Now, if we impose that $\delta{F_{pbh}}\leq{\delta{F_{back}}}$,
we obtain a constraint on the amplitude $\xi_{0}$

$$\xi_{0}\leq{(1+z_{ini})}^{2}{4\pi{\mu_{*}}^{2}
\over{3L_{0}M_{Haw}}}{\delta{F_{back}}
\over{I(\epsilon_{\odot},z_{f})}}\eqno(22)$$

Although the actual limits may be considered for any small interval, we have defined 
in practice an average background value for the radiation over a large portion of the 
spectrum as

$$<\delta
F_{back}>={\biggl[{\delta{F_{back}}\over{{10}^{-6}erg{s}^{-1}{cm}
^{-2}}}\biggr]}. \eqno(23)$$ 

Some care has yet to be taken to take into account the full
dependence of the mass density in PBHs $\varrho_{pbh}(z)$ inside
the horizon volume. After some algebric manipulation and using the
properties of the delta function we derive the expression

$$\varrho_{pbh}(z)={\biggl({1+z\over{1+z_{f}}}\biggr)}^{3}
{\xi_{0}{(M_{Haw})}^{2}\over{V_{hor}(z_{f})}}F(\mu_{*},z){\mu_{*}}
\eqno(24)$$

which is valid for any $z\geq{z_{evap}(\mu_{*})}$. If we divide
eq.(24) by $\varrho_{c}(z)=\varrho_{c}(0){(1+z)}^{3}$ (again
for $q_{0}=1/2$), we will obtain $\Omega_{pbh}(z)$ for any
$z$ which satisfies the causal relation (that is, $\mu_{pbh} \, \leq \, \mu_{hor}$),

$$(1+z_{f})\leq{4.24\times{10}^{28}\over{\sqrt{(\mu_{pbh})}}}
\eqno(25)$$

This expression precludes the existence of PBHs with masses larger
than the one contained within the cosmic horizon, i.e.
$\mu_{pbh}(z)\leq{\mu_{hor}(z)}$, but admits the case 
$\mu_{pbh} \, \ll \, \mu_{hor}$ as advocated in Refs.[6,7]. 
For $\mu_{pbh}\sim{10^{11}}$,
the critical mass value today, we have an upper limit at which
PBHs can be formed still satisfying eq.(25), namely ${z_{f}}_{max}
\sim{{10}^{16}}$.

To estimate actual bounds we evaluate the expression at $z=0$.
Then, we have

$$\varrho_{pbh} = {\xi_{0}{(M_{Haw})}^{2}{\mu_{*}}F(\mu_{*},z_{ini},z=0)
\over{{(1+z_{f})}^{3}V_{hor}(z_{f})}}\eqno(26)$$

Now, using eq.(23) and dividing it by the critical density
$\varrho_{c}(z=0)\sim2\times{10}^{-29}g{cm}^{-3}$, we have
(imposing $z_{ini}=z_{f}$)

$$\Omega_{pbh}(0)\leq{\Omega_{pbh}(z_{f},\mu_{*})}
=8.7\times{10}^{-8}{(1+z_{f})}^{2}{\mu_{*}}^{3}F(\mu_{*},z_{f},z=0)
{\epsilon_{\odot}}<\delta F_{back}>\eqno(27)$$

where $F(\mu_{*},z_{f},z=0)\sim{O(1)}$.

The interpretation of eq.(27) is the following: the right hand
side sets the maximal abundance in PBHs today ($(z=0)$) allowed by
the sky brightness assumed to be fully produced by the evaporation
of PBHs formed at $z_{f}$ with initial masses given by $\mu_{*}$. These
PBHs were born above the Hawking mass and therefore they are 
evaporating today. We have also required that their masses were initially below
the critical mass value today $\sim{10}^{26}g$, otherwise our
constraints do not apply in this form (see below). From now, we shall
denote the product $\epsilon_{\odot}<\delta F_{back}>$ which
appears conspicuously by $\epsilon_{B}$.

Our bounds based on the sky brightness will be tighter (by
construction) inside a finite region in the parameter space
defined by $\mu_{*}$ versus $z_{f}$. This finite region
corresponding to $\Omega_{pbh}(z_{f},\mu_{*})\leq{10}^{-10}$ may
be further explored to understand which is the physics allowing
these bounds. Taking the logarithm on both sides of eq.(27) we find

$$log \mu_{*} + {2\over{3}} \, log(1+z_{f})\leq{-1-{1\over{3}} \,
log{\epsilon_{B}}}\eqno(28)$$

with $\epsilon_{B} \, \equiv \, \epsilon_{\odot} <\delta
F_{back}>$ as stated above. 
This inequality displays a relation between the
initial mass $\mu_{*}$ and formation redshift $z_{f}$ bounding the
" coolest PBH population", since any PBH population below this
line will be contributing so much to the sky brightness that we 
can certainly rule it
out . This curve is a consequence of the Doppler effect, since for
$z_{f}$ above this curve, the cosmic expansion diminished the
total flux below the background value and big masses correspond to
very cold PBHs, for which the constraint is necessarily 
very poor. It is clear
that the $z_{f}$-threshold is mass-dependent and therefore, the
curve must drop at later times.

In terms of the mass we may write eq.(28) in the form

$$\mu_{*}\leq{0.1
\over{{(1+z_{f})}^{2/3}{\epsilon_{B}}^{1/3}}}\eqno(29)$$

Meaning that all those PBHs formed at $z_{f}$ satisfying both
$\mu_{*}\geq{1}$ and the eq.(28) above are much more scarce than
$\Omega_{pbh}(\mu_{*})\leq{10^{-10}}$, for a given combination value of 
the $z_{f}$ and the parameter $\epsilon_{B}$.

The very fact that $1\leq\mu_{*}\leq{\mu_{c}(z_{f})}$ leads to some
additional constraints avoiding the imposition of an $\epsilon_{\odot}$. 
Actually, since
$\mu_{c}(z_{f})=\mu_{c}(0){(1+z_{f})}^{-1}$, we note that these
relations can hold only for $(1+z_{f})\leq{10^{11}}$. But the
condition  $\mu_{*}>{1}$, eq.(28) requires also that
${\epsilon_{\odot}}{<\delta F_{back}>}\leq{{10}^{-3}
{(1+z_{f})}^{-2}}$. Taking the logarithm as before

$$-{1\over{3}}log \epsilon_{B} \geq{1+{2\over{3}}log(1+z_{f})}\eqno(30)$$.

Using that $(1+z_{f})\leq{10^{11}}$ we obtain an upper bound to
the value of the parameter $\epsilon_{B}$

$$\epsilon_{B} \leq {{10}^{-25}}\eqno(31)$$

Eq. (31) is a direct consequence of imposing $\Omega_{pbh} (\mu_{\ast}, z_{f}) \, 
\leq \, 10^{-10}$ for $\mu_{\ast} \, < \, 1$ and $z_{f} \, \gg \, 1$. We see that 
$\epsilon_{B}$ is bounded from above by a very small number, in fact so small that 
its associated distance $D_{L}(\epsilon_{\odot}) \, \sim \, 1 \, cm$. This means that we 
can exclude PBHs in the discussed conditions regardless of the irrelevant value of 
$\epsilon_{\odot}$, which is a "physical" zero.

Analogously, we must satisfy $\mu_{*}<\mu_{c}(z=0)$, which
implies $\epsilon_{B}>{10}^{-59}$.
Fig.3 displays the region between $log z_{f}$ and $log
\epsilon_{B}$ in which all those PBHs satisfying
$1\leq{\mu_{*}}\leq{0.1\over{{(1+z_{f})}^{2/3}{(\epsilon_{B})}^{1/3}}}$
are more scarce than $\Omega_{pbh}(\mu_{*})\leq{10^{-10}}$.


The region in the parameter space $log z_{f}$ versus
$log \mu_{*}$ where $\Omega_{pbh}(\mu_{*})<{10}^{-10}$ is then
bounded by the following inequalities

$$2 \, log(1+z_{f})+log \mu_{*} < \, 57.2\eqno(32)$$

$$log \mu_{*} < log\mu_{c} (0) \sim{11}\eqno(33)$$

$$log z_{f} > log z_{min} \sim{4}\eqno(34)$$

$$log \mu_{*} >{0}\eqno(35)$$

and eq.(28). It is clear that eq.(32) describes the causality
constraint, i.e. there can be no PBH with mass greater than the
horizon mass at $z_{f}$. The eq.(33) states that our method 
applies to evaporating PBHs only. Eq.(34) says that PBHs do not form as late
as $z < {10}^{4}$, since we know that the environment must be very
smooth at late times (this arbitrary prescription can be relaxed
straightforwardly). Finally, eq.(35) describes the choice
$\mu_{*}>1$. The compact region where the abundance of PBHs is
$\leq \, 10^{-10}$ in this plane is displayed in Fig.4.


\subsection{Subcase \, $\mu_{c}(z_{f})\leq{\mu_{*}}\leq{\mu_{hor}(z_{f})}$}

The range of masses implies $(1+z_{f})\leq{4.24\times{10}^{28}}$,
therefore we can rewrite the previous expressions to obtain for $\xi_{0}$

$${\xi_{0}}\leq{6.8\times{10}^{52}{(1+z_{cross}(\mu_{*}))}^{2}
{\mu_{*}}^{2}{h_{0}}^{-2}\epsilon_{B}}\eqno(36)$$

And considering that ${(1+z_{cross}\mu_{*})}^{2}
\sim{{10}^{22}{\mu_{*}}^{-2}}$ we insert these relations into
eqs.(24) and (31) to yield

$$\Omega_{pbh}(z=0,\mu_{*}\geq{\mu_{c}(z_{f})})
\leq{8.7\times{10}^{14}\mu_{*}\epsilon_{B}
F(\mu_{*},z=0,z_{ini})}\eqno(37)$$

Requiring that $\Omega_{pbh}(\mu_{*})<{10}^{-10}$ as before, the
corresponding masses must satisfy

$$\mu_{c}(z_{f})\leq\mu_{*}\leq{{10}^{-25}\over{\epsilon_{B}}}
\eqno(38)$$

Analogously to the former case $\mu_{*}>{1}$ then the cutoff is limited to

$$\epsilon_{B}\leq{{10}^{-25}}, \eqno(39)$$

And the same considerations as before apply.
Since $\mu_{*} < \mu_{c}(z=0)\sim{10^{11}}$ must also hold, we obtain the limit

$$\epsilon_{B} \geq {10}^{-34} , \eqno(40)$$

again corresponding to microscopic lengthscales, and therefore not 
relevant for the bounds so obtained. 
Finally, requiring that ${10^{-25}\over{\epsilon_{B}}}\geq{\mu_{c}(z_{f})}$, 
we obtain the inequality

$$log{\epsilon_{B}\over{(1+z_{f})}}\leq{-36} \eqno(41)$$

Note that eqs.(38) and (39) are completely independent of $z_{ini}$ because 
the latter dependence
cancels out. Then, it is eq.(39) the one that 
sets the maximum value for the lower cutoff in
redshift, already shown to be irrelevant.
Although we may have considered the subcase $\mu_{*}>\mu_{c}(z_{f})$ in two parts,
(the first $1 > \mu_{c}(z_{f})$ and the second 
$1<\mu_{c}(z_{f})$), they actually differ only in the allowed value of
$z_{f}$: for the first we have $(1+z_{f})>{10}^{11}$ and for the
second case, $(1+z_{f})<{10}^{11}$.

Figure 5 displays a region in the plane $log(1+z_{f})$ versus $\epsilon_{B}$ in 
which we find PBHs
satisfying $\Omega_{pbh}(\mu_{*})\leq{10^{-10}}$ and
for which $\mu_{*}\geq{\mu_{c}(z_{f})}$. All these PBHs satisfy 
the conditions
$\mu_{*}\geq{1}$, $\mu_{*}\leq{\mu_{c}(0)}\sim{10}^{11}$ and $log{z_{f}}\leq{28.6}$ as well
(for $log{z_{f}}\geq{28.6}$ we would have $\mu_{hor}\leq{1}$).


\bigskip

\centerline{\bf Second Regime: $\mu_{*} < 1$}

\bigskip

When we deal with PBHs born with masses smaller than the Hawking
mass initially, we may substitute the lower limit of the integral
eq.(21) by $z_{evap}(\mu_{*})$, because no PBH will survive its
own timescale for evaporation. Therefore the integral
$I(z_{evap}(\mu_{*}),z_{ini})=\int_{z_{evap}(\mu_{*})}^{z_{ini}}
dz{(1+z)\over{D_{L}^{2}(z)}}$ is explicitly evaluated as

$$I(z_{evap},z_{ini})={(H_{0}/c)}^{2}\biggl[G(\mu_{*})+
ln{\biggl({z_{ini}\over{z_{evap}(\mu_{*})}}
\biggr)}\biggr] \eqno(42)$$

where $G(\mu_{*})={z_{ini}-z_{evap}(\mu_{*})\over{z_{evap}(\mu_{*})z_{ini}}}$ 
and $z_{evap}(\mu_{*})={1\over{\mu^{2}}} - 1$.

Following the same reasoning as in the previous case we obtain for $\xi_{0}$

$$\xi_{0}\leq{7.2\times{10}^{22}{h_{0}}^{-2}{<\delta F_{back}>
\over{G(\mu_{*})+ln(z_{ini}{\mu_{*}}^{2})}}} \, {g}^{-1} \eqno(43)$$

where $z_{ini}$ falls in one of these two subcases: when
$\mu_{*}>\mu_{c}(z_{f})$ we have $z_{ini}={10^{11}/\mu_{*}}$ and
when $\mu_{*}<\mu_{c}(z_{f})$ we have $z_{ini}=z_{f}$.

Using the expressions above we finally find, after substituting
eq.(43) into eq.(31) and dividing by $\varrho_{c}(z)$

$$\Omega_{pbh}(z>z_{evap}(\mu_{*})) < {9\times{10}^{14}}F(\mu_{*},z_{ini},z)
{\biggl[{\mu_{*}\over{G(\mu_{*})+ln(z_{ini}{\mu_{*}}^{2})}}\biggr]}
<\delta F_{back}>{h_{0}}^{-2}
\eqno(44)$$

Since that $F(\mu_{*},z_{ini},z=z_{evap}(\mu_{*}))=0$, the result eq.(44) describes
correctly the evolution of the abundance from $z_{f}$ until
$z_{evap}(\mu_{*})$, taking into account the Hawking evaporation. The most
interesting case obtained from eq.(44) applies to $z=z_{f}$, from which
we derive the maximum abundance allowed at formation still consistent
with the sky brightness today $<\delta F_{back}>$)

$$\Omega_{pbh}(z_{f})<9\times{10}^{14}\biggl[{\mu_{*}\over{G(\mu_{*})+
ln(z_{f}{\mu_{*}}^{2})}}
\biggr]<\delta F_{back}>{h_{0}}^{-2}\eqno(45)$$

for $\mu_{*}\leq{\mu_{c}(z_{f})}$ and

$$\Omega_{pbh}(z_{f})\leq{9\times{10}^{14}}
\biggl[{\mu_{*}\over{G(\mu_{*})+ln(10^{11}\mu_{*})}}\biggr]
<\delta F_{back}>{h_{0}}^{-2}\eqno(46)$$

for $\mu_{*}\geq{\mu_{c}(z_{f})}$.

However, when we try to force eqs.(45)
and (46) to yield values $\Omega_{pbh} \leq 10^{-10}$
we did not find physical windows for reasonable values of
PBH masses. The reason is
that to obtain $\Omega_{pbh} \, \leq \, 10^{-10}$; $\mu_{\ast}$ have 
to be very small and thus 
violate our semi-classical assumptions. This limitation is to be expected,
since those PBHs evaporated a long time ago when the cosmic
radiation was $\sim \, {(1+z)}^{4}$ times more intense than the present
background. This means that we can not use the cosmic isotropic
radiation in order to estimate constraints for the sub-Hawking PBH
population, and other methods (as described, for example, by Green
et al. \cite{[LLL99]}) must be used to obtain useful limits to their
abundance.

\bigskip

\section{\bf Conclusions}

\bigskip

We have shown that under a set of circumstances a limit on $\Omega_{pbh}$ better
than ${10^{-10}}$ can be obtained by using the background brightness of the sky.
This is quite stringent
if we compare it with the values given by other methods \cite{[LG98]}. 
We attempted to
evaluate the evolution of maximal abundance $\Omega_{pbh}(z)$ for any $z$
from observations performed
today ($z=0$), a process that requires integration over $z$ (and eventually
over the actual mass spectrum from whatever the formation process, not attempted here).

The good news here is that PBHs can be excluded in a fairly 
large window (as seen in Fig. 4) if they were born above the Hawking
mass (but below $10^{26} \, g$). We see four physical reasons for this fact:

1) Many of these PBHs {\it still} evaporate today, and therefore contribute to the 
background in various bands. 

2) The cosmic radiation today
is much less intense than at high redshift values, and therefore
constitutes a useful natural background. 

3) The cosmic expansion
today is quite weak, then the Doppler damping of the emitted
radiation does not have any killing effect. 

4) PBHs initially
above the critical mass at some $z_{f}$ delay the start of their evaporation,
therefore they injected energy later, when the expansion is less
effective to damp the radiation.

The constraints so obtained are weakly
dependent of $q_{0}$ or the cosmological constant, 
and the mass range probed by 
this method is larger than
previous works. Even though we take into account
the existence of the critical mass and the
redshift of the emitted radiation through the
function $I(\epsilon_{\odot},z_{f})$, we found that our 
analysis renders useful limits for PBH masses initially 
above $\sim 10^{15} \, g$, but not for those below it. In addition, PBHs
above ${10}^{26}g$ must be limited by other
methods because they do not evaporate at all. 
The consideration of the different cases leaded us to conclude that 
$\epsilon_{\odot}$ need not to be imposed by hand. Moreover, it happens 
to be an extremely small number in all cases, with microscopic associated 
distances $D_{L}(\epsilon_{\odot})$. Physically, this means that the 
spatial distribution of PBHs does not need to satisfy any specific 
requirement near the earth, and thus the bounds so obtained are quite general. 

The upper limits to the number density of PBHs were obtained from the simple
requirement
$\delta{F}_{pbh} \, < \, \delta{F}_{back}$, since other
astrophysical mechanisms may be also contributing to the cosmic background brightness
as measured by our detectors. The limits have been derived for a Dirac's
delta mass function, whose main features are
much simpler than any other choice. We can
think this case as a first approach to the general problem, likely accurate
for peaky distributions of a more general type. Detailed
calculations of the latter remain an interesting problem for future investigations.

\section{\bf Acknowledgments}

We would like to thank D.M\"uller and C.R.Ghezzi for valuable discussions
and technical support with numerical recipes that were used in this work.
P.S.Custodio has been supported by Funda\c c\~ao de Amparo \`a Pesquisa do
Estado de S\~ao Paulo (FAPESP) and J.E.Horvath acknowledges partial financial
support from CNPq (Brazil).

\vfill\eject
\noindent
{\bf Figure captions}

\bigskip
\noindent
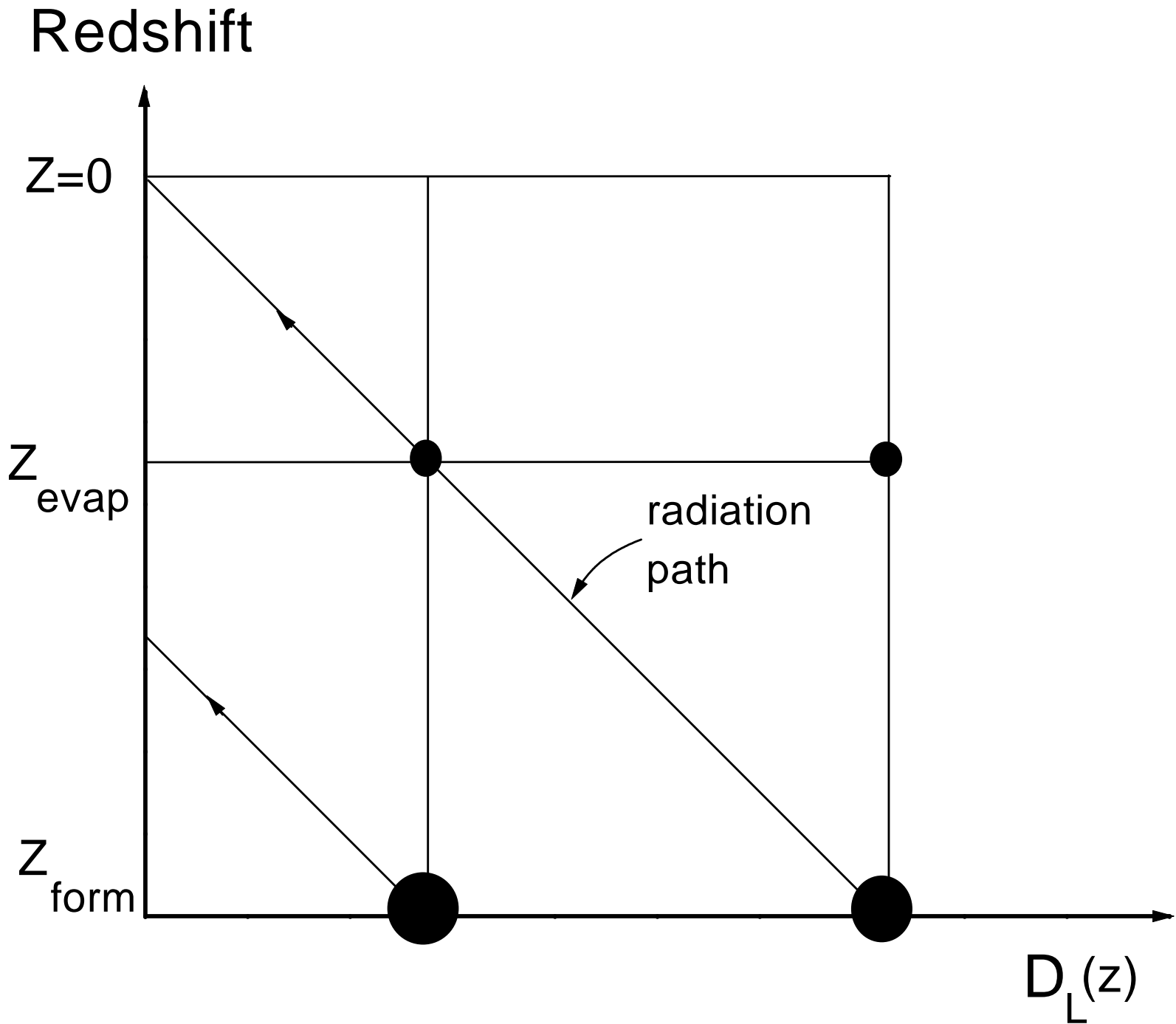
Figure 1. The fate of primordial black holes in the redshift vs. distance-luminosity plane.
A PBH formed at $z_{form}$ evaporates and the emitted radiation arrives at the observer when its path (
explicitely indicated) crosses the vertical axis. If the PBH mass is initially below the Hawking mass, then  , the corresponding redshift for complete evaporation $z_{evap}$ is located at $z \geq 0$. 

\bigskip
\noindent
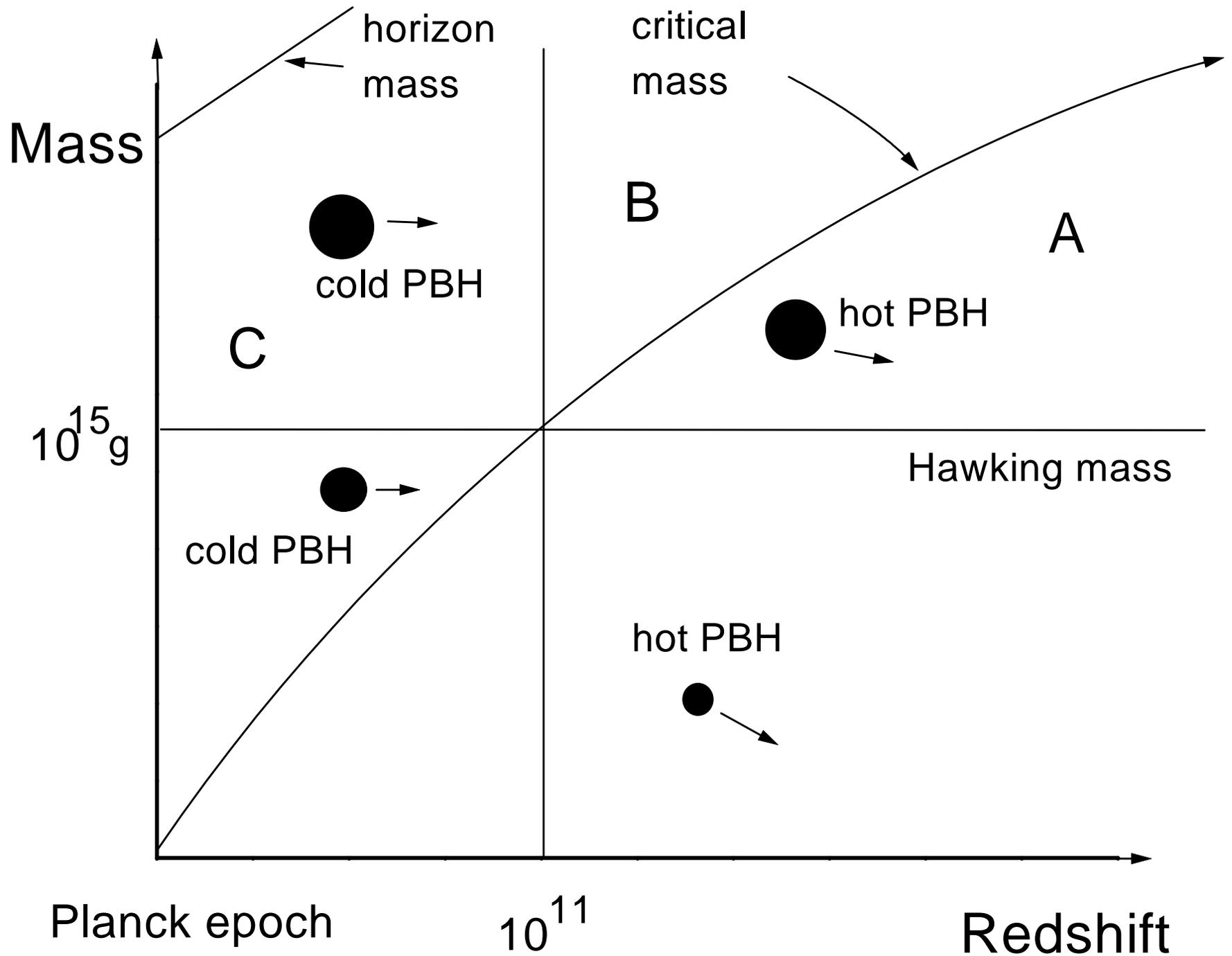
Figure 2. Mass scales vs. redshift of formation plane. The critical mass curve and the Hawking mass constant are explicitely indicated. Note that both cross at $z \sim 10^{11}$. 
In the region A hot (evaporating) PBHs begun to evaporate right after their 
formation. In the region B the PBHs did not begin to evaporate at formation, but only after crossing the critical mass curve and entering into the region A.
In the region C PBHs were formed at very high redshift, greater than $z \, = \, 10^{11}$ and 
followed a similar history. Below $M_{Haw} \sim 10^{15} \, g$, PBHs that evaporated completely today. Our method of analysis and bounds thus obtained are useful for the PBHs formed at the regions A, B and C (termed "First Regime" in the text) but not below the Hawking mass (see text).

\bigskip
\noindent
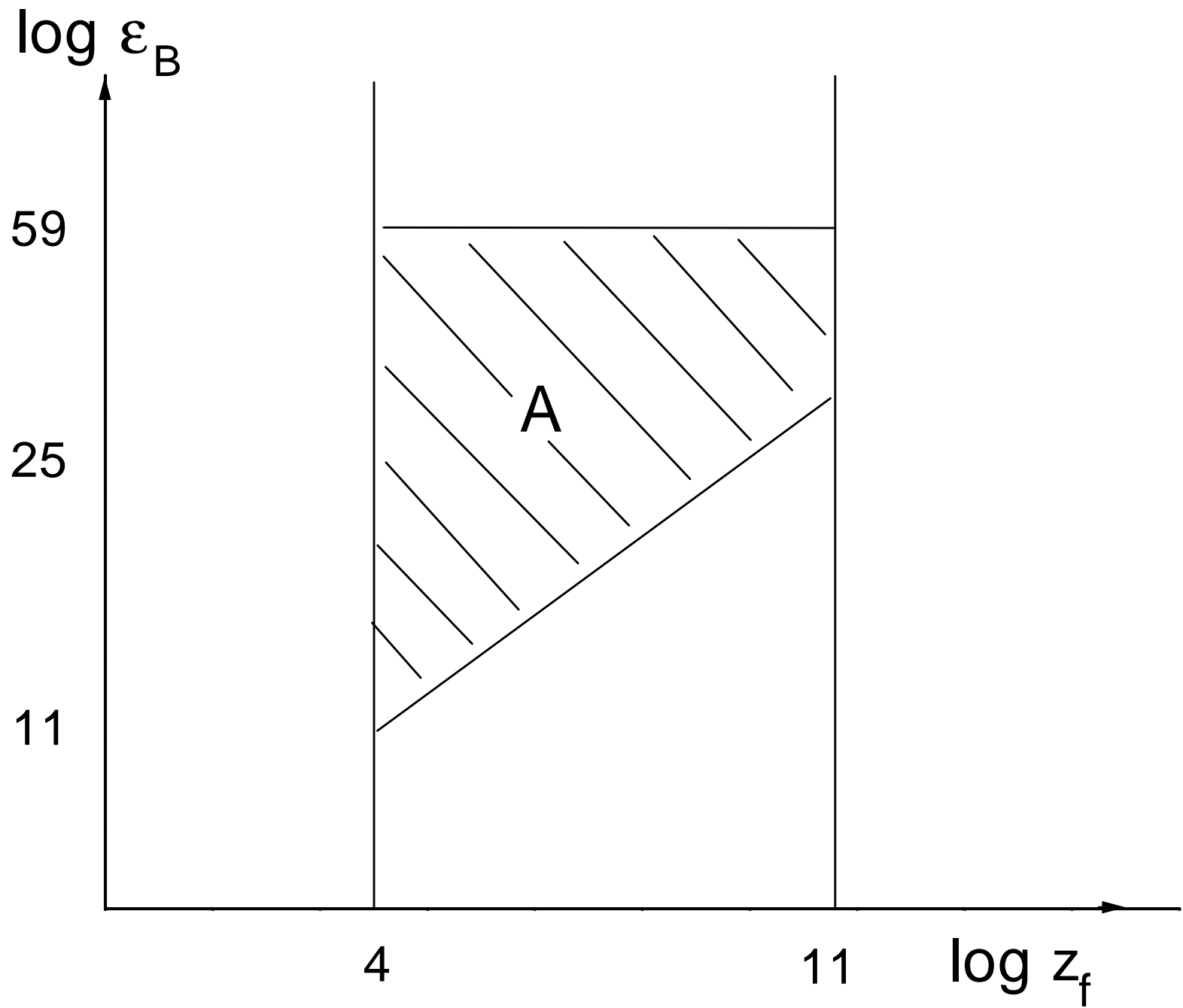
Figure 3. The region inside which $\Omega_{pbh} \, < \, 10^{-10}$ in the 
$\epsilon_{B}$ vs. $z_{f}$ (hatched) for those PBHs born in region A. The boundaries of this region are defined in the text. 

\bigskip
\noindent
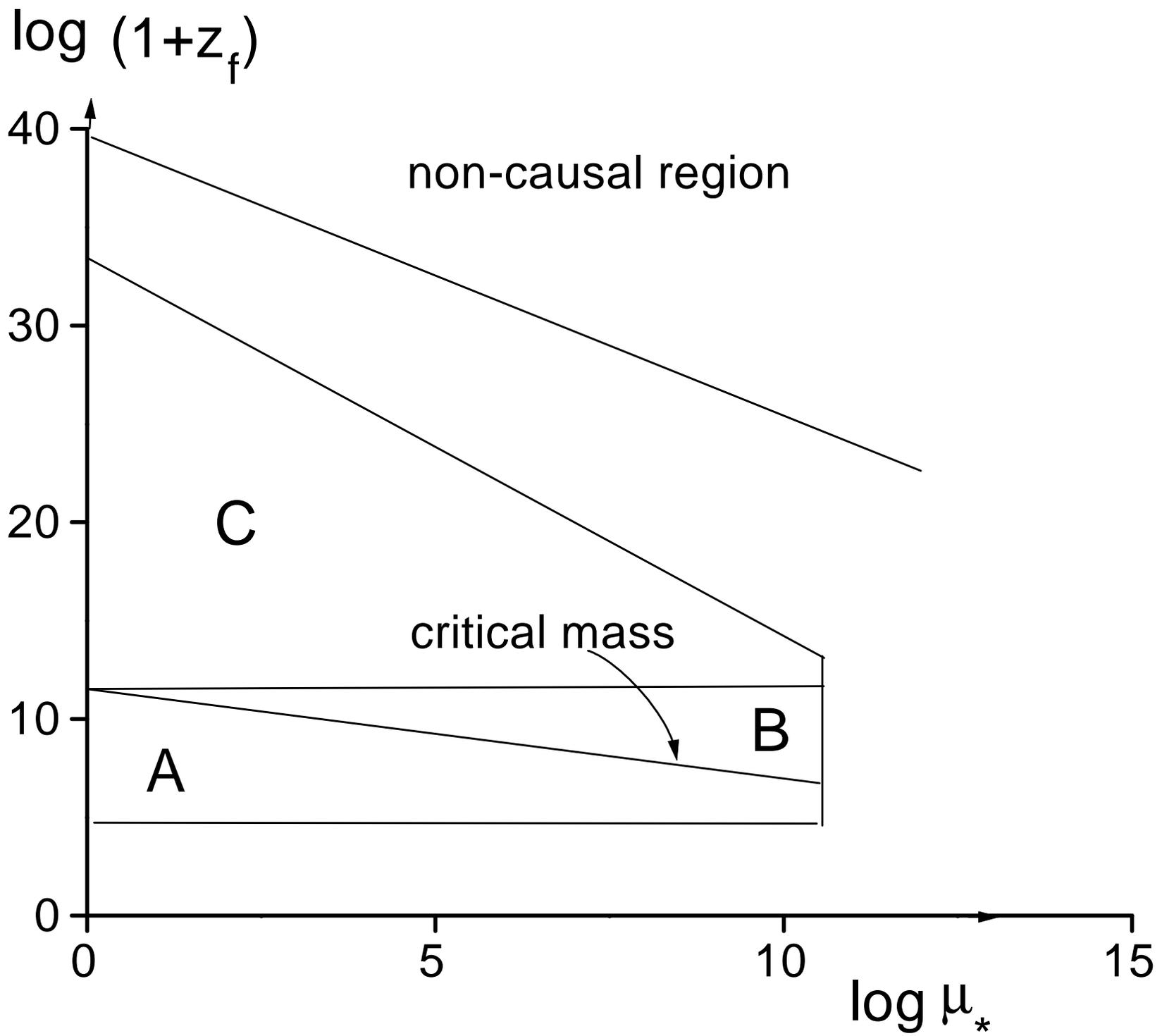
Figure 4. The compact region in which the $\Omega_{pbh} \, < \, 10^{-10}$ 
in the redshift vs. mass plane (see text). At $log \mu_{\ast} \sim 11$ the PBHs have a mass of $10^{26} \, g$, the critical mass value today. 
The inclined straight line bounding region C from above is set by the Doppler effect of the cosmological expansion. We have also drawn the horizon mass line 
and explicitly indicated the non-causal region above it. The regions A, B and C 
correspond to the regimes and subcases defined in the text.

\bigskip
\noindent
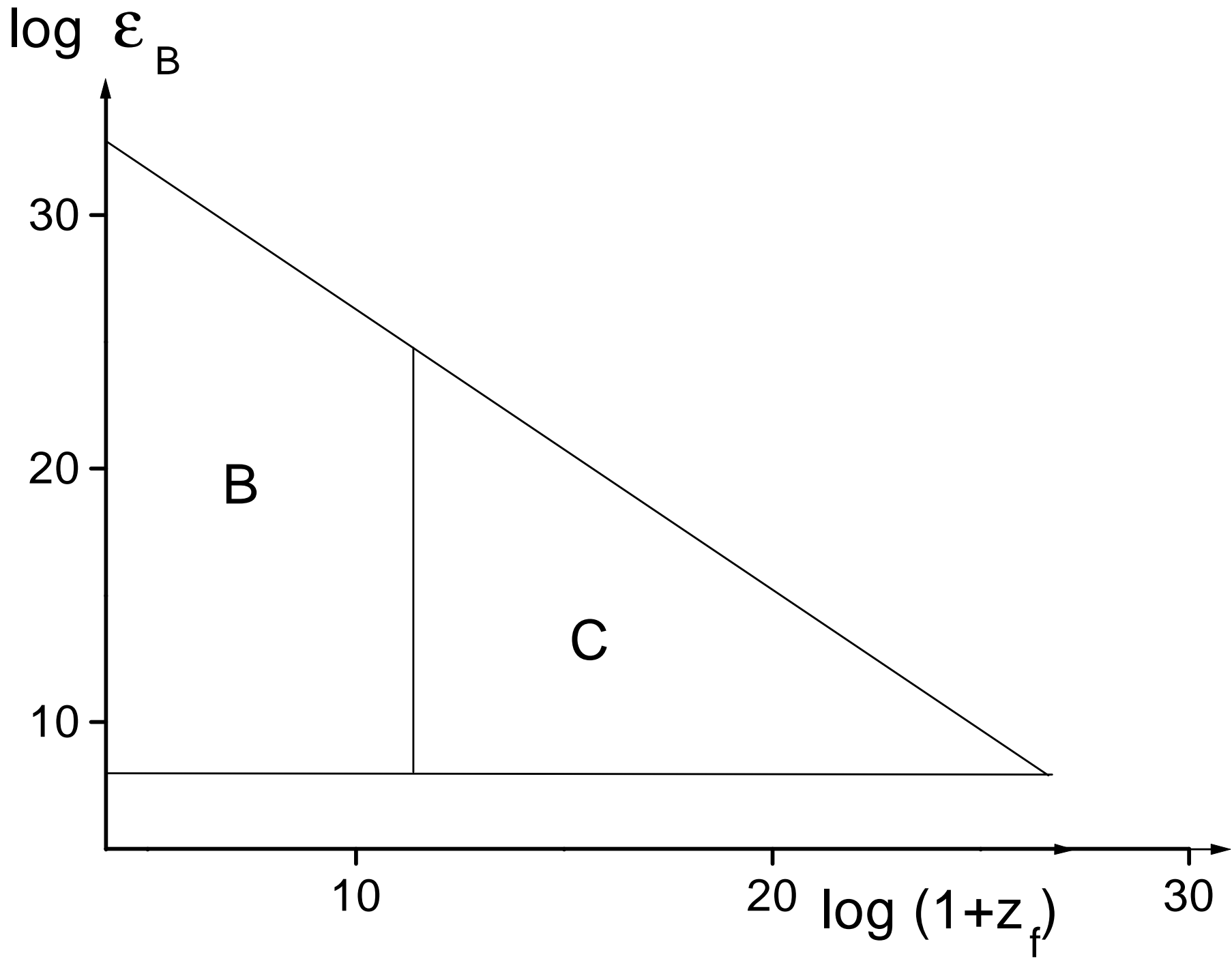
Figure 5. The same as Fig. 3 for the regimes corresponding to the regions B and C. The region B spans redshifts from $0$ up to $10^{11}$, when the critical mass equals the Hawking mass.
The region C corresponds to any redshift value bigger than $10^{11}$ up 
to $10^{28}$, the moment when the horizon mass equals the Hawking mass (of academic interest only).
The upper diagonal line reflects the combination of the Doppler effect and the value 
of $\epsilon_{B}$ (see text). 
   
\end{document}